\documentclass[floatfix,amsmath,amssymb,aps,prb,groupedaddress,twocolumn,superscriptaddress]{revtex4-2}
\usepackage[margin=1.7cm]{geometry}

\usepackage[dvipsnames]{xcolor}
\usepackage{graphicx}
\usepackage{amsthm,amssymb,amsmath,braket,mathdots,mathtools}
\usepackage{bm}

\usepackage{psfrag}
\usepackage{relsize,amsbsy}
\usepackage[export]{adjustbox}
\usepackage{dsfont}
\usepackage{tikz}

\usepackage[utf8]{inputenc}
\definecolor{mypink}{RGB}{255,105,180}

\usepackage{mwe}
\usepackage{tabularx}
\usepackage{multirow}
\usepackage[normalem]{ulem}
\usepackage[colorlinks,bookmarks=true,citecolor=blue,linkcolor=red,urlcolor=blue]{hyperref}
\usepackage[capitalize]{cleveref}
\usepackage{algorithm}
\usepackage{algpseudocode}

\newcommand{\norm}[1]{\lVert #1 \rVert}

\newcommand{\1}{\mathds{1}}

\newcommand{\matr}[1]{\mathsf #1}
\newcommand{\op}[1]{#1}
\newcommand{\Nw}{\tilde{N}}

\newcommand{\bit}{\begin{enumerate}}
\newcommand{\eit}{\end{enumerate}}

\definecolor{MDG}{rgb}{0,0.55,0.05} 

\begin{document}
\title{Efficient tensor network simulation of multi-emitter non-Markovian systems}

\author{Irene Papaefstathiou}
\affiliation{Max-Planck-Institut f{\"u}r Quantenoptik, Hans-Kopfermann-Str. 1, D-85748 Garching, Germany}
\affiliation{Munich Center for Quantum Science and Technology (MCQST), 80799 Munich, Germany}
\author{Daniel Malz}
\affiliation{Department of Mathematical Sciences, University of Copenhagen, Universitetsparken 5, 2100 Copenhagen, Denmark}
\author{J. Ignacio Cirac}
\affiliation{Max-Planck-Institut f{\"u}r Quantenoptik, Hans-Kopfermann-Str. 1, D-85748 Garching, Germany}
\affiliation{Munich Center for Quantum Science and Technology (MCQST), 80799 Munich, Germany}
\author{Mari Carmen Ba\~nuls}
\affiliation{Max-Planck-Institut f{\"u}r Quantenoptik, Hans-Kopfermann-Str. 1, D-85748 Garching, Germany}
\affiliation{Munich Center for Quantum Science and Technology (MCQST), 80799 Munich, Germany}

\date{\today}

\begin{abstract}
   We present a numerical method to simulate a system of multiple emitters coupled to a non-interacting bath, in any parameter regime.
    Our method relies on a Block Lanczos transformation that maps the whole system onto a strip geometry, whose width is given by the number of emitters.
    Utilizing the spatial symmetries of the problem and identifying the relevant range of energies of the bath we achieve a more efficient simulation, which we perform using tensor network techniques.
   As a demonstration, we study the collective emission from multiple emitters coupled to a square lattice of bosons and observe how the departure from Markovianity as coupling strength and emitter number is increased prevents collective radiation. We also simulate the dynamic preparation of an excitation in a bound state from a multi-excitation initial state.
    Our work opens new possibilities for the systematic exploration of non-Markovian effects in the dynamics and equilibrium properties of multi-emitter systems. 
    Furthermore, it can easily be extended to other setups, including finite bath temperature or impurities coupled to fermionic environments.
\end{abstract}

\maketitle
\section{Introduction}\label{sec:Introduction}

The paradigm of open quantum systems is ubiquitous: realistic quantum systems are coupled to an environment that causes decoherence and dissipation~\cite{breuer2007}.
When the system--bath coupling is weak and the bath can be assumed to be Markovian, a Lindblad master equation can be derived, which allows one to simulate the system without explicit reference to the bath dynamics~\cite{breuer2007}.
Many realistic systems do not fulfil these assumptions and the bath dynamics has to be explicitly accounted for. This makes non-Markovian systems computationally challenging to treat, and consequently there is an ongoing research effort to develop methods to simulate the diverse families of non-Markovian systems~\cite{Vega2017},
which has led to a number of techniques to study non-Markovian dynamics of open quantum systems~\cite{tanimura1989, makri1995, makri1995meth, garraway1997, Degenfeld2014, smith2014, schroder2017multi, Strathearn2017EfficientNQ,Pollock2018, link2024}.

A particular important family of open quantum systems are those in which the environment can be taken to be \emph{non-interacting}. This family includes systems studied in quantum optics where the system(s) are atoms and the bath non-interacting photons~\cite{Cohen-Tannoudji1998}, quantum impurity models in condensed matter (where the bath are non-interacting fermions)~\cite{Hewson1993,Anderson1961} and models of spins coupled to non-interacting phonons, e.g., in biomolecules~\cite{Thorwart2009}, or solid-state emitters~\cite{Doherty2013}.

In pioneering work, Wilson introduced chain-mapping techniques to study impurity problems~\cite{Wilson1975}.
Following a suitable discretization of the bath modes, the degrees of freedom of the non-interacting bath could be mapped to a chain and treated numerically with the iterative numerical renormalization group (NRG) algorithm, now of widespread use for equilibrium impurity problems~\cite{Bulla2008}. NRG produces a ground state solution in the form of a matrix product state (MPS)~\cite{Verstraete2008,Schollwoeck2011}. 
The more general tensor network picture~\cite{Verstraete2008} allowed unifying NRG with the variational density matrix renormalization group (DMRG)~\cite{White1992} and adopt 
MPS time evolution techniques~\cite{Vidal2004,Daley2004,White2004}
to study non-equilibrium impurity problems~\cite{Weichselbaum2009}. In quantum optics settings, combining MPS techniques with a discretization of the bath using orthogonal polynomials~\cite{Prior2010,Chin2010} has been used in multiple problems, including, for instance, problems in photochemistry~\cite{Chin2013nat},
cavity QED in the ultrastrong coupling regime~\cite{SanchezMunoz2018} or emitters coupled strongly to a waveguide~\cite{noachtar2022}. The strategy can be generalized to treat baths at finite temperature, too~\cite{devega2015thermofield,Schwarz2018}.
A key advantage of the method in these setups is that the error it incurs is controlled and can in principle be lowered arbitrarily by expending more computing power~\cite{Woods2015,Vega2015,Trivedi2021}. While very successful for single impurities, extending these techniques to many impurities with a shared bath has been a challenge~\cite{Bulla2008}.
This makes it difficult to study the bath-mediated interaction of a collection of impurities beyond the well-studied case of two impurities~\cite{Jayaprakash1981}.

In quantum optics, we are often interested in systems comprising many emitters coupled to a bosonic (photonic) bath, which underlies a wealth of collective effects such as super- and subradiance~\cite{Gross1982}, bound states~\cite{Shi2016}, and collective enhancement in light-matter coupling~\cite{Chang2014}, as used for example for quantum memory~\cite{Lvovsky2009a}. These phenomena are fairly well-understood in the Markovian setting, but due to the complexity and limitations of numerical techniques, much less is known in the non-Markovian setting.

Recent experimental advances in controlling the photonic environment of quantum emitters through nano-photonic structures~\cite{Chang2018,Sheremet2023} makes it possible to study the dynamics of quantum emitters in \emph{structured} baths, which have been predicted to exhibit a host of unconventional effects~\cite{GonzalezTudela2024}.
These include fundamentally non-Markovian effects that arise due to slow light~\cite{Baba2008} or strong coupling~\cite{FornDiaz2019} such as the modification of superradiance in the presence of retardation~\cite{Sinha2020}, new or modified quantum information protocols~\cite{GonzalezBallestero2013,Pichler2015,Ramos2016}, or nonperturbative effects in structured baths~\cite{GonzalezTudela2017,GonzalezTudela2017a}.
There now exist several experimental platforms that can be used to systematically explore dynamics of emitters strongly coupled to structured baths, such as circuit electrodynamics~\cite{liu2017, sundaresan2019, Andersson2019, Ferreira2020} or neutral atoms~\cite{stewart2020}.

These developments motivate the search for universal methods to simulate multi-emitter (or multi-impurity / multi-orbital) systems coupled to non-interacting baths.
In principle, the mapping of a single quantum system coupled to a bath into a chain can be generalized to many quantum systems coupled to the same bath~\cite{BLShirakawa,allerdt2015,noachtar2022} by using several seed vectors in the Block Lanczos recursive technique \cite{GOLUB1977}.
Instead of a chain, the resulting system becomes a strip, its width equal to the number of systems coupled to the bath.
This strip can be collapsed back into a ``fat'' chain, where each site now has a Hilbert space dimension exponential in the width (and thus in the number of emitters or impurities)~\cite{BLShirakawa}, which however makes the simulation inefficient.

Here, we also employ the Block Lanczos technique to map multi-impurity or multi-emitter problems to a strip geometry, but propose key improvements that make it numerically tractable to study real time evolution in such systems.
To demonstrate the enhanced Block Lanczos algorithm, we numerically study the out-of-equilibrium, many-excitation, non-Markovian dynamics of many emitters coupled to a two-dimensional structured reservoir of non-interacting bosons.
By truncating the Block Lanczos basis, we can efficiently study effectively infinite systems for sufficiently long times as to access physical phenomena, such as collective superradiant emission in the non-Markovian setting, and dynamical bound state formation from a highly excited initial state.
We additionally exploit the point group symmetry of the bath to obtain a more efficient representation, which allows us to increase the number of emitters studied with this method. An additional truncation in the energy levels of the bath allows us to decrease the Trotter step error and consequently extend the times we can simulate.
With these improvements, we simulate systems of up to six emitters with as many excitations as emitters in both the weak-coupling (weakly non-Markovian) and strong-coupling (strongly non-Markovian) regime using tensor network techniques~\cite{Verstraete2008,SCHOLLWOCK201196,Orus2014annphys,Silvi2019tn,Okunishi2022,Banuls2023}.

The structure of the paper is as follows: In Sec.~\ref{sec:Methods} we introduce the $\textit{symmetry-enhanced Block Lanczos algorithm}$ we use to study the time evolution of a general system that is coupled to a bosonic, non-interacting bath, in Sec.~\ref{sec:Model} we present the model of interest and in Sec.~\ref{sec:diagonal} and in Sec.~\ref{sec:diamond} we discuss the results and the details of our numerical implementation. We close with a discussion in Sec.~\ref{sec: Discussion}.

\section{Symmetry-enhanced BL algorithm}\label{sec:Methods}

In this section of the paper, after introducing the general setting, we proceed by presenting the $\textit{symmetry-enhanced Block Lanczos algorithm}$ we use to numerically study out-of-equilibrium dynamics. We base our approach on the Block Lanczos technique, already explored before, to map the 
initial system-bath Hamiltonian to a (quasi) one-dimensional system. We improve the efficiency of the algorithm by utilising the symmetries of the two-dimensional bath and by identifying and using only the range of bath energies which are the most relevant for the time evolution. These steps allow us to study more challenging setups. We finally present the numerical method for simulating time evolution with the use of tensor network techniques. 

\subsection{Setting}
\label{sec: setting}

Let us consider a system coupled to a bosonic bath evolving under the Hamiltonian
\begin{align}
	H &= H_S + H_B + H_\mathrm{int},
    \label{eq: Hamiltonian operators}
\end{align}
where ${H}_{S}$ the Hamiltonian of the system, ${H}_{B}$ the Hamiltonian of a non-interacting, bosonic bath and ${H}_{\mathrm{int}}$ describes the interaction between the system and the bath, which has to be linear in the bath modes.

The bath and interaction Hamiltonians can thus be written as

\begin{align}
    {H}_{B}&=\sum_{m,n=1}^{L_{B}}a_n^\dagger [\matr H_B]_{nm} a_m = \sum_{q=1}^{L_{B}}\epsilon_q\tilde a_q^\dagger \tilde a_q  \nonumber \\
	{H}_{\mathrm{int}}&= \sum_{i=1}^{N_{O}}\sum_{n=1}^{L_B} \left( g^{(i)}_n O^{(i)}a^{\dagger}_n +\mathrm{h.c.} \right).
	\label{eq: single body terms}
\end{align}

Here, $a_{n}$($a_{n}^\dagger$) are the annihilation(creation) bosonic bath operators, $\tilde a_q$ ($\tilde a_q^{\dagger}$) are the corresponding diagonal modes, with corresponding energies $\epsilon_q$, $O^{(i)}$ are operators acting only on the system, and $g_n^{(i)}$ are coupling strengths. Notice that there is no assumption on the form of the operators $O^{(i)}$. The number of modes in the bath is $L_{B}$ and $N_{O}$ is the total number of system operators.
Here and in the rest of the manuscript we denote operators in the Hilbert space using normal letters (e.g., $\op H_S$) and matrices using sans font (e.g., $\matr H_B$).

The aim is to describe quasi-exactly the dynamics of the system coupled to a non-interacting bath. For this, we map the Hamiltonian $\op{H}$ to a quasi one-dimensional system, which allows us to employ numerical Matrix Product States (MPS) techniques~\cite{Verstraete2008,SCHOLLWOCK201196} to study the time-evolution of the system. 

In \cref{sec: Block Lanczos} we describe how to transform the bath Hamiltonian $H_{B}$ with the use of a Block Lanczos transformation.
For this transformation the bath Hamiltonian needs to be non-interacting, but there is no restriction on the geometry or the dimensionality of the bath. Note that even though we are considering a bosonic bath, the block diagonalization proceeds in the same way for a fermionic bath.

\subsection{Block Lanczos}
\label{sec: Block Lanczos}

\begin{figure*}[t!]
\centering
    \includegraphics[width=0.95\linewidth]{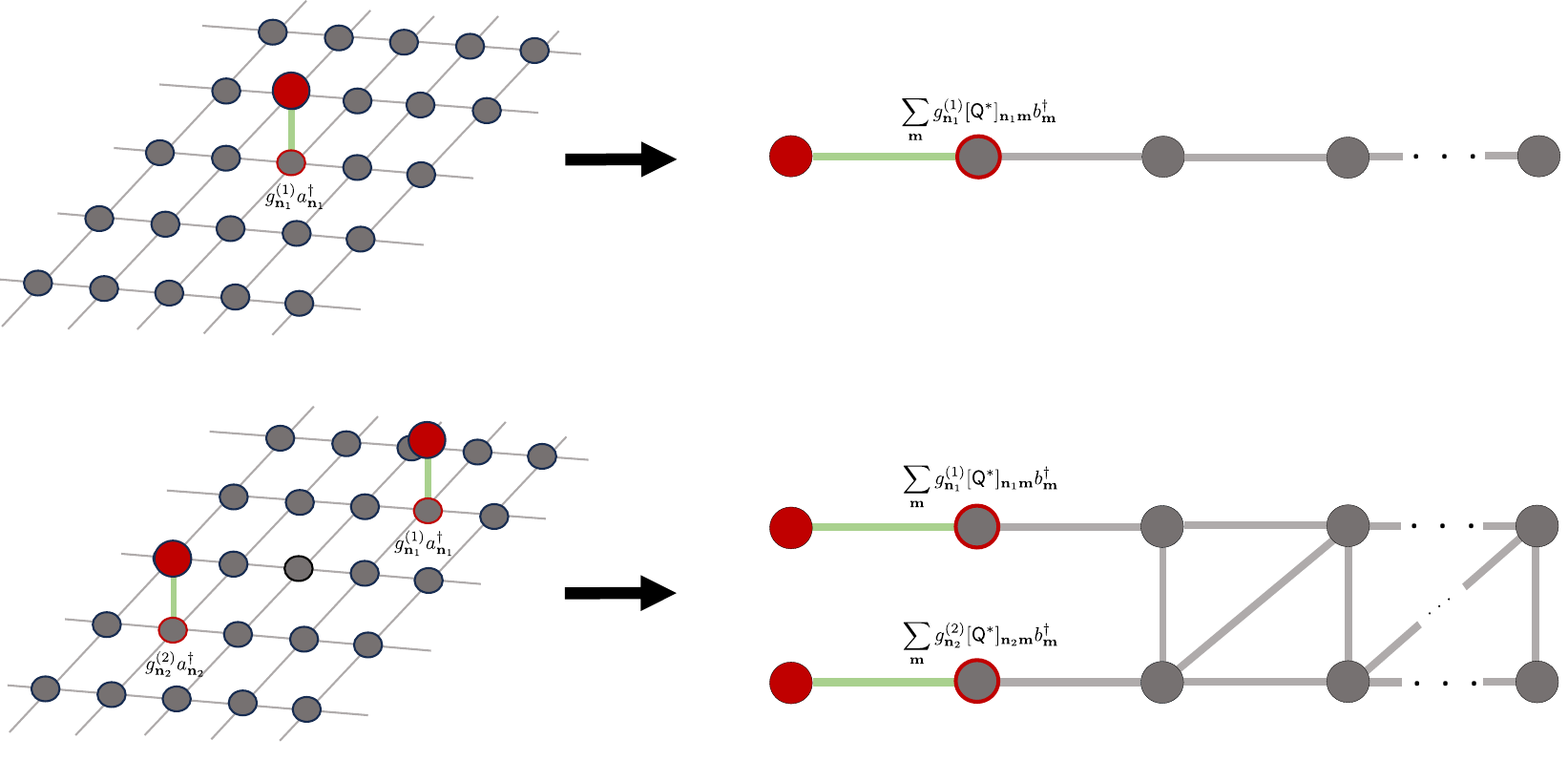}
\caption{Block Lanczos transformation for a set of quantum systems (for single emitter above, for two emitters below) coupled locally to a two-dimensional bath on a square lattice. The result is a ladder of width equal to the number of emitters, with the emitter sites (red circles) on the edge. The interaction Hamiltonian directly couples (green lines) each emitter to a single bath mode, which acts as seed for the BL transformation (grey circles with red outline).  
The BL transformation produces the ladder geometry, with Hamiltonian terms connecting pairs of bath modes (grey lines).
We note here that the structure of the resulting system only depends on the number of system operators and does not depend on the structure or dimensions of the bath.}
\label{fig: mappingSquareLattice}
\end{figure*}

The original Lanczos algorithm~\cite{lanczos1950iteration}, introduced to find extreme eigenvalues of Hermitian matrices, provides a way to tridiagonalize a matrix by iteratively constructing a suitable basis. This is done by iteratively applying the matrix to a given initial (\emph{seed}) vector $q$, and orthogonalizing the resulting set of vectors, a process which yields a basis in the Krylov subspace $\mathcal{K}(\matr A,q,n)=\mathrm{span}\{ q, \matr Aq, \matr A^{2}q, \dots, \matr A^{n-1}q \}$. Written in this basis, the original matrix becomes tridiagonal.

The Block Lanczos (BL) method~\cite{cullumBL,GOLUB1977} is a generalization of the original Lanczos algorithm, which, instead of a single seed vector, introduces a block of $N_O$ of them $\{q^{(i)}\}$, for $i=1,\dots,N_O$.
Iteratively applying the matrix to all of the seed vectors and orthogonalizing defines a basis in the space $\bigcup_i \mathcal K(\matr A,q^{(i)},N_O)$, and in this basis the original matrix becomes block-tridiagonal.
In the context of quantum many-body problems, the BL method was first used in Ref.~\cite{shirakawa2014block} to address quantum impurities. Here we follow the same approach to render the Hamiltonian $\op{H}$ treatable with MPS techniques.
 
We apply the Block Lanczos process to the bath Hamiltonian $\matr H_B$ using as seed vectors the bath modes the system couples to, that is, we choose as initial block a set of $N_O$ vectors ${q}^{(i)}$ with components $q^{(i)}_n\propto g^{(i)}_n$ (up to normalization) [see Eq.~\ref{eq: single body terms}].
We assume here that $q^{(i)}$ are  orthonormal. This is the case, for instance, if each system operator couples to a single distinct bath mode.
If this is not the case, the initial $q^{(i)}$ are simply obtained by orthonormalizing the vectors given by the $g^{(i)}_n$ couplings.
We then collect the seed vectors into one $L_B\times N_O$ matrix $\matr{Q}_{1}=[q^{(1)}, q^{(2)}, \dots, q^{(N_O)}]$, which constitutes the first block of the BL basis.

Defining $\matr{E}_{i}=\matr{Q}_{i}^{\dagger}\matr H_{B}\matr{Q}_{i}$ we can then construct the rest of the Block Lanczos basis by finding $\matr Q_{i}$ through the recursion 
\begin{align}
    \matr{R}_{i}&=\matr{H}_{B}\matr{Q}_{i}-\matr{Q}_{i}\matr{E}_{i}-\matr{Q}_{i-1}\matr{T}_{i-1}
    \label{eq: Block Lanczos recursion}
\end{align}
with $\matr{Q}_{0}=0$, $\matr{T}_{0}=0$.
Subsequent $\matr{Q}$ and $\matr{T}$ matrices are obtained from the QR factorization $\matr{R}_{i}=\matr{Q}_{i+1}\matr{T}_{i}^{\dagger}$.
Therefore the matrices $\matr{T}_{i} \in \mathbb{C}^{N_O \times N_O}$ generated by the recursion have lower triangular form, which means that $(\matr{T}_{i})^{m'm}=0$ for $m' < m$. 

To enhance clarity we present the first step of the recursion
\begin{align}
\matr{R}_{1}&=\matr{H}_{B}\matr{Q}_{1}-\matr{Q}_{1}\matr{E}_{1}=(\mathbb{I}-\matr{Q}_{1}\matr{Q}_{1}^{\dagger})\matr{H}_{B}\matr{Q}_{1}
\end{align}

Since $\matr{Q}_{1}\matr{Q}_{1}^{\dagger}$ is the projector onto the subspace of seed modes, by performing a QR decomposition of $\matr{R}_{1}$ we identify a new block of vectors, orthogonal to the seed modes, to which $\matr{Q}_{1}$ gets directly coupled by $\matr{H}_{B}$. This is the second block of the Block Lanczos basis, $\matr{Q}_{2}$.

We repeat the recursion of Eq.~\eqref{eq: Block Lanczos recursion} $L$ times, obtaining the $L_B\times N_O$ matrices $\matr{Q}_{i}$ and constructing the matrix $\matr{Q}$ as
\begin{align}
    \matr{Q}=\Big(\matr{Q}_{1}, \matr{Q}_{2}, \dots, \matr{Q}_{L}    \Big)
\end{align}
where $\matr{Q}^{\dagger}\matr{Q}=\mathbb{I}$.
Note that if $L=L_B/N_O$, then $\matr Q$ is unitary.

Using $\matr Q$, we can define the block-tridiagonal BL Hamiltonian

\begin{align}
   \matr{H}^{BL}_{B}&=\matr{Q}^{\dagger}\matr{H}_{B}\matr{Q}
   =
	\begin{pmatrix}
    	\matr{E}_{1} & \matr{T}_{1} & 0 & 0 \cdots \\
    	\matr{T}_{1}^{\dagger} & \matr{E}_{2} & \matr{T}_{2} & 0 \cdots \\
    	0 & \matr{T}_{2}^{\dagger} & \matr{E}_{3} & \matr{T}_{3}  \cdots \\
    	\vdots & \vdots & \vdots & \vdots  \ddots
	\end{pmatrix}
 \label{eq:Hbl}
\end{align}
We highlight here that the Block Lanczos technique was utilized in order to transform only the basis of the bath operators.
This basis transformation results in a quasi-one-dimensional ladder geometry (see Fig.~\ref{fig: mappingSquareLattice})
where the width is given by the number of system--bath coupling operators $(N_O)$ and the length is given by the number of iterations $L$. 
Since $\matr T_i$ have triangular form, the Hamiltonian terms connect only modes within the same rung or in neighboring ones.
\begin{algorithm}[H]
\caption{Block Lanczos Algorithm}
\begin{algorithmic}[1]
\Statex \textbf{Input:}  Matrix $\matr H_{B} \in \mathbb{C}^{L_{B} \times L_{B}}$, $\matr H_{B}^{\dagger}=\matr H_{B}$, starting matrix $U \in \mathbb{C}^{L_{B} \times N_{O}}$, maximum number of iterations $L_a$
\State $\matr{Q}_0 \gets 0$
\State $\matr{T}_0 \gets 0$
\State $\matr{Q}_{1} \gets U$
\For{$i \gets 1$; $i \le L_{a}$}
    \State $L \gets i$ 
    \Comment{Current number of blocks}
    \State $\beta_i \gets \|\matr{Q}_{i}\|_F$
    \If{$\beta_i = 0$}
        \State \textbf{break}
    \EndIf
    \State $\matr{E}_i \gets \matr{Q}_{i}^{\dagger}\matr{H}_{B}\matr{Q}_{i}$
    \State $\matr{R}_{i} \gets \matr{H}_{B}\matr{Q}_{i}-\matr{Q}_{i}\matr{E}_{i}-\matr{Q}_{i-1}\matr{T}_{i-1}$
    \State $[\matr{Q}_{i+1}, \matr{T}_{i}^{\dagger}] \gets \text{QR}(\matr{R}_{i})$ \Comment{QR decomposition of $\matr{R}_{i}$}
\EndFor
\Statex \textbf{Output:} Orthonormal Basis $\matr{Q} \in \mathbb{C}^{L_{B} \times N_{O}L}$
\end{algorithmic}
\label{alg:BL}
\end{algorithm}

Under this transformation the bath operators become
\begin{align}
    b_{n}&=\sum_{m} \matr{Q}_{nm}^{\dagger}a_{m}
\end{align}
with the bath Hamiltonian now written as
\begin{align}
    H_{B}&=  \sum_{m,n=1}^{L_{B}}b_{m}^{\dagger}[\matr{H}^{BL}_{B}]_{mn}b_{n}
\end{align}
whereas the interaction Hamiltonian becomes
\begin{align}
    {H}_{\mathrm{int}}&= \sum_{i=1}^{N_{O}}\sum_{n=1}^{L_B}\sum_{m=1}^{L_B} \left( g^{(i)}_n O^{(i)}[\matr{Q}^{*}]_{nm}b^{\dagger}_m +\mathrm{h.c.} \right).
\end{align}

\begin{figure*}[t!]
\centering
    \includegraphics[width=0.90\linewidth]{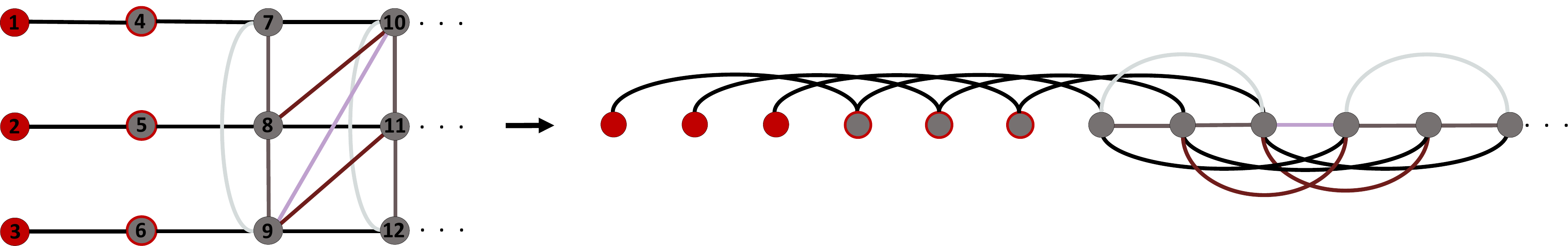}
    \caption{
	The ladder structure produced after the Block Lanczos transformation (left) and its mapping to an 1D system with long range interactions (right), for the case of three emitters. The lattice sites of the emitters are denoted by red circles (all $\textit{legs}$ in the first $\textit{block}$), and located at the beginning of the ladder or the 1D system. The  gray circles represent the lattice sites of the (bosonic) chain. The gray circles with the red outline are the bath modes that are coupled to the emitters. They are selected by the choice of the input seed vectors of the Block Lanczos algorithm. The hoppings are denoted by lines. As shown here, the Hamiltonian of Eq.~\eqref{eq: Hamiltonian operators} can be mapped onto a one-dimensional model, at the expense of long-range interactions. The maximum interaction range depends on the number of system operators, $N_{O}$, as well as the ordering of the operators. As shown in this figure, we choose a particular ordering, for which the maximum range of interactions is $N_{O}$.}
\label{mapping1Dchain}
\end{figure*}

This geometry is particularly suitable to be described by a tensor network ansatz.
In particular, to use MPS, we map the ladder to a chain (see Fig.~\ref{mapping1Dchain}), which results in terms of range up to $N_O$. 
Thinner strips are substantially easier to simulate, which puts some limitations on the number of distinct points the system can couple to the bath (or, equivalently, the number of emitters).
The length of the strip can be chosen according to the desired maximum simulation time $t$.
Formally, the BL transformation is exact if we take $L=L_B/N_O$ such that $\matr Q$ is unitary. 
But since the coupling along the length of the strip is nearest neighbour, transport along the strip is subject to some maximum propagation speed, which introduces a light cone.
Therefore, given a maximum simulation time $t=t_{\mathrm{trunc}}$, we can stop the BL iteration after some $L_{\mathrm{trunc}}<L_B/N_O$ such that the light cone is covered.
This argument could be formalized in terms of Lieb-Robinson bounds, which can be derived in a variety of ways, e.g., by noting that the total particle number is bounded (which applies in our case)~\cite{Schuch2011a} or even for a bounded energy density as long as the nearest-neighbour interaction is some form of hopping (which also applies in our case)~\cite{Yin2022}.
The BL Hamiltonian $\matr H_B^{BL}$ is then truncated to ($N_{O}L_{\mathrm{trunc}})\times (N_{O}L_{\mathrm{trunc}}$).

\subsection{Symmetries}
\label{sec: symmetries}

The ladder structure makes the problem more amenable to tensor network methods, but the presence of long $N_O$-range terms limits the practical application to problems with a reduced number of system operators. Nevertheless, we can exploit the symmetries of the problem to reduce the width of the strip, which can substantially reduce numerical cost.
In particular, we will consider the discrete symmetries of the bath lattice, and, for concreteness, discuss the case of a two-dimensional square lattice, although the scheme is easily generalizable to other geometries.

\begin{figure*}[t]
\centering
 \includegraphics[width=0.95\linewidth]{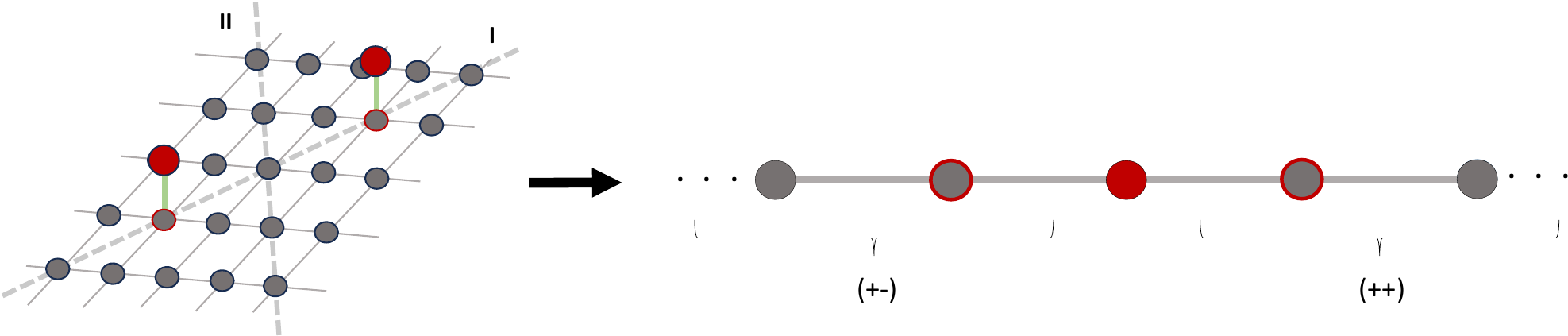}
\caption{The figure shows the chain produced for the diagonal configuration of two emitters, after utilising the symmetry-enhanced Block Lanczos algorithm. The red circles are the emitters and the gray the lattice sites of the bosonic chain that are either symmetric with respect to both diagonal I and diagonal II (++), or symmetric with respect to diagonal I and antisymmetric with respect to diagonal II (+-).}
\label{fig:symmetric_chain}

\end{figure*}
Let us consider a discrete symmetry operation $U_{\alpha}$ (e.g. a spatial reflection or rotation) that commutes with the bath Hamiltonian $[\op{H}_{B},U_{\alpha}]=0$. 
Because $\op{H}_{B}$ is block-diagonal with respect to the symmetry sectors, 
the Block Lanczos transformation can be applied independently in each of them.
The coupling of the system to the bath can also be written as a sum over sectors, as each of the initial vectors $q^{(i)}$ introduced in \ref{sec: Block Lanczos} is a direct sum of components in different symmetry sectors.
Using these components as initial vectors for the Block Lanczos procedure in each sector, we obtain as many ladder structures, coupled through the emitters, in a \emph{star} geometry. Whereas each Block Lanczos transformation is now run on a smaller dimensional problem, in the most general case, the width of each of the ladders is still $N_O$.

A further optimization is possible when the total Hamiltonian preserves (part of) the spatial symmetries of the bath. 
For concreteness, we consider a square bath lattice 
of size $L \times L$ with $L$ odd, and periodic boundary conditions, and we will assume that the system operators couple to individual spatial bath modes.
The center of the lattice is the point $(0,0)$, and  the symmetry group is the dihedral group, $D_{4}$. The non-Abelian $D_{4}$ group is generated by one reflection and one $\pi /2$ rotation and has 5 irreducible representations (irreps).
We can construct bases of spatial modes that transform according to these irreps.
However, the different configurations of the emitters partially break this symmetry, and it is more convenient for our mapping to choose only two sectors that correspond to the remaining symmetry, while ensuring a quasi-one-dimensional geometry.
More concretely, we will use spatial reflections with respect to two orthogonal axes $\mathcal{D}_{1}$ and $\mathcal{D}_{2}$, chosen according to the setup (see fig.~\ref{fig:symmetric_chain}).
The total Hamiltonian $\op H$ is invariant under these reflections, and we can label four symmetry sectors according to the symmetric or antisymmetric character with respect to each of them, as $(s_1,s_2)$, with $s_{1,2}=\pm$. 
For a generic point in the lattice $(x,y)$, these reflections define an \emph{orbit} of four modes
$a_{\pm x,\pm y}$, out of which we can construct four modes with well-defined $(s_1,s_2)$. Modes lying along the $\mathcal{D}_{1,2}$ axes, or the mode at the origin have smaller orbits. Thus each sector contains approximately one fourth of the bath modes.

Written in terms of the symmetric bath modes, the interaction term identifies linear combinations of system operators that couple to specific sectors, and makes explicit how different sectors may couple to each other through the interaction with the emitters. 
We consider emitter configurations and couplings that also respect the above symmetry operations,
hence each single bath mode with well-defined symmetry will only couple to a linear combination of the emitters whose positions correspond to the same orbit of spatial points.
We can group these multiplet of emitters together in a new effective system, and use the bath symmetric modes as seeds for the BL in each sector.
Since each such mode couples to a single system (containing multiple emitters), the width of the ladder is now reduced to the number of such multiplets.
Even though the dimension of the effective systems is now larger, MPS methods are capable of dealing with sites with relatively large local dimension, and the reduction in the width by a constant factor allows us to multiply the number of emitters that we can treat by the same number.

\begin{figure}[t!]
\centering
    \includegraphics[width=0.90\linewidth]{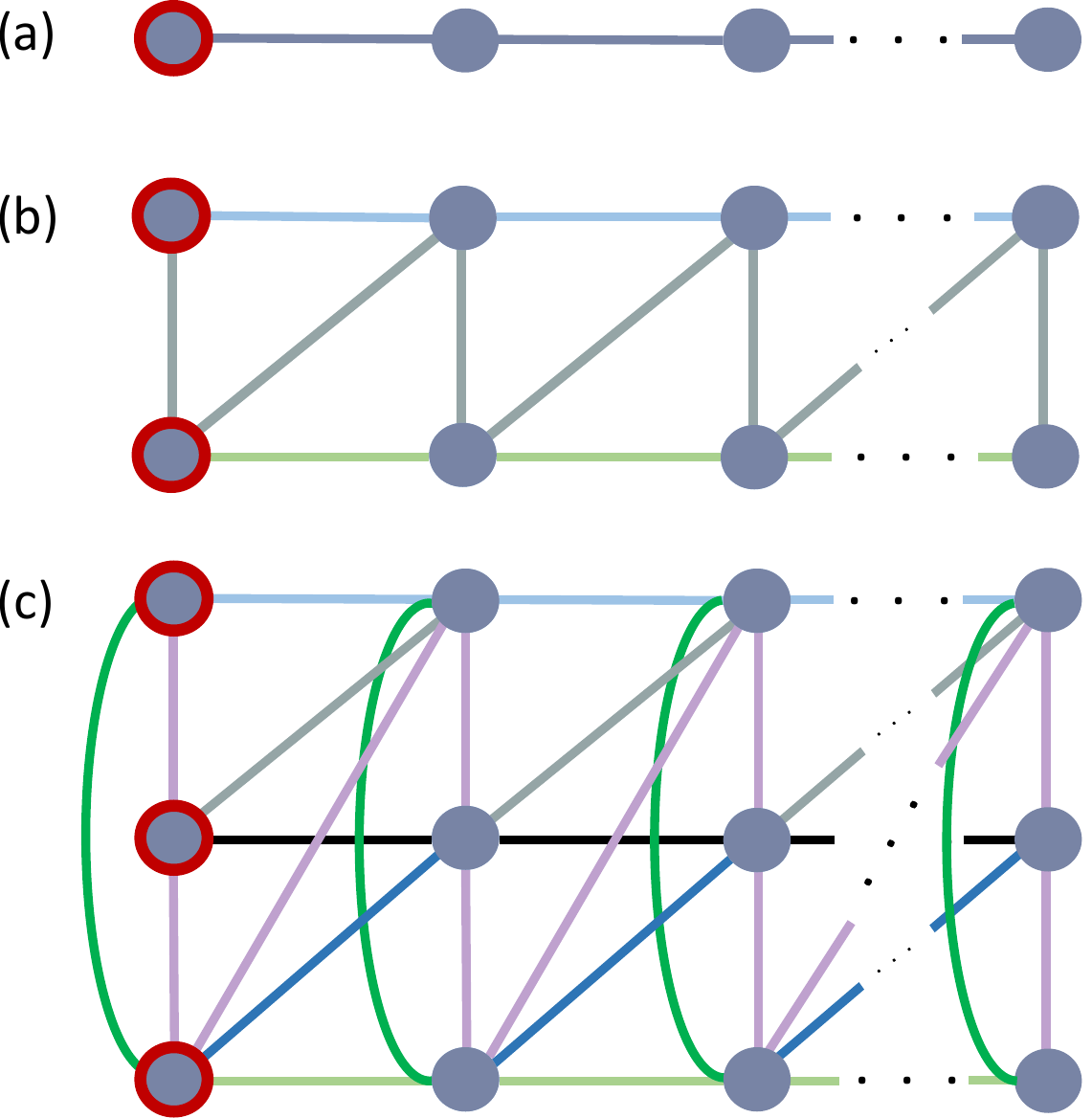}
\caption{
    We illustrate how we group the terms of $H$ into sets whose exponential terms are aplpied together as a single MPO, as part of the numerical method presented in Sec.~\ref{sec: numerical method}. Here the system operators are omitted, with all the blue circles representing the bath modes. The first block (blue circles with red outline) represents the bath modes that are coupled to the system  (seed modes).
    Note that in contrast to \cref{fig: mappingSquareLattice,mapping1Dchain}, the seed modes a slightly different, because we have truncated the bath in energy space, which results in non-zero couplings to multiple bath operators. 
    Lines of the same color indicate terms that are applied together in a single MPO.
     The number $M$ of such MPOs depends on the number of operators, $N_{O}$. For $N_{O}=1$, as seen in Fig.~\ref{fig: choice of exponentials}(a), there is only a single Hamiltonian, whereas $M=4$ (Fig.~\ref{fig: choice of exponentials}(b)) and $M=8$ (Fig.~\ref{fig: choice of exponentials}(c)) for $N_{O}=2$ and $N_{O}=3$ correspondingly.}   
\label{fig: choice of exponentials}
\end{figure}

\subsection{Time evolution with truncated energy levels}
\label{sec: energy truncation}
Under the Lanczos mapping, the hopping along the chain is upper bounded by the spectral norm $\norm{\matr H_B}_2$ of the bath Hamiltonian [\emph{cf.}\ \cref{eq: single body terms}].
Consequently, a bath Hamiltonian with a large spectral range as compared to the system--bath coupling typically transforms into a chain with large hopping rates compared to the characteristic system time scale.
As a result, we have to use long chains and small time steps (compared to the natural system time scale) in the numerical integration, which makes the algorithm inefficient.
This is a common setting in quantum optics, where we have emitters at a certain frequency ($\omega_0$, say) weakly coupled to electromagnetic modes with a large frequency range.

\begin{figure}[t]
\centering
    \includegraphics[width=0.95\linewidth]{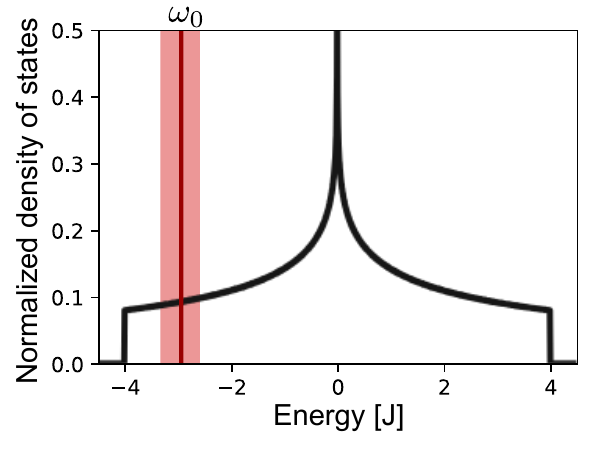}
\caption{Density of states of a two-dimensional tight binding model with hopping rate $J$. If we simulate emitters at frequency $\omega_0$ coupled to this bath at some coupling strength $g_0$, it is sufficient to consider bath modes close in frequency to this scale as explained in \cref{sec: energy truncation}.}
\label{fig:dos}
\end{figure}

To make this more precise, suppose that the system Hamiltonian $H_S$ has discrete eigenstates with energies that are (approximately) multiples of one (or potentially several) underlying frequencies $\omega_0$ and that the system operators $O^{(i)}$ correspond to transitions between these states. For example in quantum optics we are often interested in systems of several emitters, each of which has a ground and an excited state that are separated by a transition frequency $\omega_0$, and that couple to a bath of electromagnetic modes via raising and lowering operators between those levels.
In this situation, we can truncate the bath also in energy space by removing bath modes that are far detuned from the transitions in the system.

Given a transition frequency $\omega_0$, we thus separate the bath into two parts: (i) the modes with energies in a window $[\omega_0-\Delta,\omega_0+\Delta]$, which we collect in the Hamiltonian $\matr H_B^{\mathrm{keep}}$ and (ii) the rest of the modes, which we collect in $\matr H_B^{\mathrm{discard}}$ such that $\matr H_B=\matr H_B^\mathrm{keep}\oplus \matr H_B^{\mathrm{discard}}$. 
For the truncated bath, we have $\norm{\matr H_B^{\mathrm{keep}}-\omega_0\1}_2\leq\Delta$. Performing a BL transformation on the truncated Hamiltonian thus yields a strip with hopping rates bounded by $\Delta$, which means that we can use shorter chains and larger Trotter steps, yielding much more efficient simulations.
We illustrate this schematically in \cref{fig:dos}.
Note that this procedure can be generalized to several transition frequencies, for which we need to use several energy windows.

If $\Delta\gg|g_{\alpha n}|$ for all $\alpha, n$, the modes in $\matr H_B^{\mathrm{discard}}$ are too far detuned to play a role, and their predominant contribution is a Lamb shift (which can be obtained from a Markov approximation) of the size
\begin{equation}
    \delta \omega_\alpha = \sum_{\epsilon_n\notin[\omega_0-\Delta,\omega_0+\Delta]}\frac{|g_{n,\alpha}|^2}{\omega_0-\epsilon_n}.
\end{equation}
Later, we will consider emitters that are coupled to a single site of an array of bath modes with some coupling rate $g_0$, which is the same for every emitter $\alpha$. In this case, $\sum_{n}|g_{n,\alpha}|^2=g_0$, such that $\delta\omega_\alpha<g_0/\Delta$.
In practise, we find the best $\Delta$ numerically by simulating the single-particle problem exactly for increasing values of $\Delta$ until convergence is reached.

\subsection{Numerical implementation}
\label{sec: numerical method}

The ladder geometry obtained after the BL transformation can be numerically simulated using TNS methods. In particular, we use a matrix product state (MPS) ansatz to describe the state of the system~\cite{Verstraete2008,SCHOLLWOCK201196}. 
To this end, we map the ladder to a one-dimensional chain by choosing a linear ordering of the sites, as illustrated in Fig.~\ref{mapping1Dchain}.
For the symmetry-enhanced transformation described in~\ref{sec: symmetries}, the emitters sites occupy a central rung in the ladder (respectively a set of contiguous sites inside the chain), with the half ladder (half chain) to each side corresponding to a different symmetry sector (see fig.~\ref{fig:symmetric_chain}). 
We write a MPS ansatz for this chain and simulate the real time evolution of the system by means of the t-MPS algorithm~\cite{paeckel2019}.
To this end, the Hamiltonian is decomposed in a sum $\op H=\sum_{p} \op{H}_p$, where each part $\op{H}_p$ contains only mutually commuting terms
and where the evolution operator for time $t$ is approximated by Trotter expansion with steps of 
length $\delta$. For instance, to the first order this reads $e^{-i\op{H}t}\approx \left(\prod_p  e^{-i\op{H}_{p}\delta}\ \right)^{t/\delta}$.
In the t-MPS algorithm, each $e^{-i\op{H}_{p}\delta}$ is written as a matrix product operator (MPO)~\cite{verstraete2004,zwolak2004,Pirvu_2010}, whose action onto the MPS state is 
approximated variationally (see e.g.~\cite{paeckel2019} for details of the algorithm).

Whereas splitting $H$ as a sum is standard for nearest-neighbour Hamiltonians, in the case of a Hamiltonian with longer range terms, 
the choice of the $\op{H}_p$ terms is more open, and can severely affect the computational cost of the algorithm.
Our model contains terms of range up to the width of the BL ladder $\Nw$, arranged with a particular geometry, corresponding to the lower triangular structure of the 
$\matr{T}$ blocks in \eqref{eq:Hbl}. 
Using a first order Trotter expansion, we split the Hamiltonian in two-body terms involving either two bath modes or one (effective) system and one bath site, 
for each of which the exponential can be computed exactly and expressed as a MPO.
Instead of applying the resulting  $O(\Nw L)$ terms one by one, which would result in a large overhead in the running time of the algorithm, we group sets of non-crossing terms 
into MPOs with the same bond dimension of one of the included gates, and apply them simultaneously using the t-MPS method.
The number of MPOs depends on the width of the ladder having $O(\Nw^2)$ as an upper bound, as we illustrate in Fig.~\ref{fig: choice of exponentials}.
One time step of evolution is achieved after applying the whole set of MPOs to the state given as a MPS.

In general, the evolution generates entanglement and an accurate description of the resulting state requires a larger bond dimension. This can limit the feasible simulations to times for which the required bond dimension can be handled by the computational resources at hand. Beyond that time, the truncation error becomes dominant.
In our setups, however, the low number of excitations in the system ensures that the generated entanglement is bounded, and we can simulate the evolution for very long times with tensors of moderate dimensions. 

An MPS describes systems with finite dimensional Hilbert spaces. In order to treat bosonic systems, some kind of truncation is required. This can be done via an optimization of the  finite dimensional basis for the bosonic modes~\cite{zhang1998boson,guo2012boson}. In our case, however, the situation is simple because the total number of excitations in the system is conserved and upper-bounded by the number of emitters, as the initial state of the bath is always empty. 
We could thus limit the occupation number of each bosonic mode to a finite value 
$n_{\max}$, equal to this initial number of excitations, without incurring in numerical error.
In practice, and in order to reduce the computational cost, we can actually use a lower upper bound $n_{\max}$ without affecting the results (as we verify checking convergence within our numerical precision).

\section{Results}

\subsection{Model} 
\label{sec:Model}

We consider a system comprising $N_e$ two-level systems (quantum emitters) with free Hamiltonian $\op{H}_E$ coupled to a non-interacting bosonic bath on a 2D square lattice of size $L_{B}= N \times N$. The free bath Hamiltonian, denoted by $H_B$, includes only hopping between the nearest-neighbouring modes.
To remove finite-size effects, we choose the size of the lattice large enough that reflections from its boundaries do not play a role. The rich physics of this model has previously been studied in the single-excitation regime in Ref.~\cite{gonzalez20172Dreservoirs, gonzalez2017markovian} and a key motivation here is to extend these results to the many-excitation regime.

We consider a Hamiltonian of the following form

\begin{align}
    H&=H_{B}+H_{E}+H_{\mathrm{int}}  \nonumber \\
    &= -J\sum_{\langle \mathbf{n},\mathbf{m} \rangle} \big(a_{\mathbf{n}}^{\dagger}a_{\mathbf{m}}+h.c.)\nonumber \\ &+ \Omega\sum_{i=1}^{N_{e}} \sigma^{+}_{i}\sigma^-_{i}+ g\sum_{i=1}^{N_{e}} \big( a_{\mathbf{n}_{i}}\sigma^{+}_{i} +h.c. \big),
\label{eq: Total Hamiltonian}
\end{align}
with nearest-neighbour hopping rate $J>0$, emitter frequency $\Omega$, emitter--bath coupling strength $g$, and emitter lowering operator $\sigma_{i}$.
In this model, the identical emitters are each coupled to a single spatial bath mode (at position $\mathbf{n}_{i}$), but of course the method here can be used to study arbitrary disordered frequencies, hopping, and coupling strengths. Here we did not include counter-rotating terms, since our underlying motivation stems from systems considered when studying the interaction of atoms with nanophotonic structures in the optical regime~\cite{Chang2018, GonzalezTudela2017}. In this regime the bandwidth of the bath, $J$, can be made small by engineering suitable nanophotonic structures. In these settings, even when $g/J$ is strong, the rotating-wave approximation is still valid. Notice, nevertheless, that the method can also be used to study Hamiltonians with counter-rotating terms, as it only requires a linear coupling term as specified in ~\cref{eq: single body terms}. 

For periodic boundary conditions, the bath Hamiltonian is diagonalized by momentum modes
$a_{\mathbf{k}}=\frac{1}{\sqrt{N}}\sum_{\textbf{n}}a_{\textbf{n}}e^{-i\textbf{k}\cdot \textbf{n}}$, for $\textbf{k}=(k_{x}, k_{y})$
and $k_{i}\in \frac{2\pi}{N}\big(-\frac{N}{2},\cdots,\frac{N}{2}-1\big)$,
\begin{align}
H_{B}&=\sum_{\mathbf{k}}\omega_{\mathbf{k}} a^{\dagger}_{\mathbf{k}}  a_{\mathbf{k}},
\label{eq: bath in momentum space}
\end{align}
with dispersion relation $\omega_{\mathbf{k}}=-2J[\cos(k_{x})+\cos(k_{y})]$.
The band covers the energy range $\omega_{\mathbf{k}}\in[-4J, 4J]$, with the density of states shown in fig.~\ref{fig:dos}. 

This model exhibits non-Markovian effects in a range of parameter regimes, where methods like Lindblad master equations and Green functions cannot be used.
Already in the single-excitation case, the give rise to exotic emission effects, as explored in Ref.~\cite{GonzalezTudela2017}. In the following, we explore the many-excitation regime of this model.

As described in Sec.~\ref{sec: energy truncation}, in order to allow for longer Trotter steps and make the simulations more efficient, we truncate the energy space of the bath, keeping energies only within an interval $[\Omega-\alpha g,\Omega+\alpha g]$. The value of the factor $\alpha$ is found numerically, such that it ensures convergence of the results.

\begin{figure}[t]
\centering
 \includegraphics[width=0.95\linewidth]{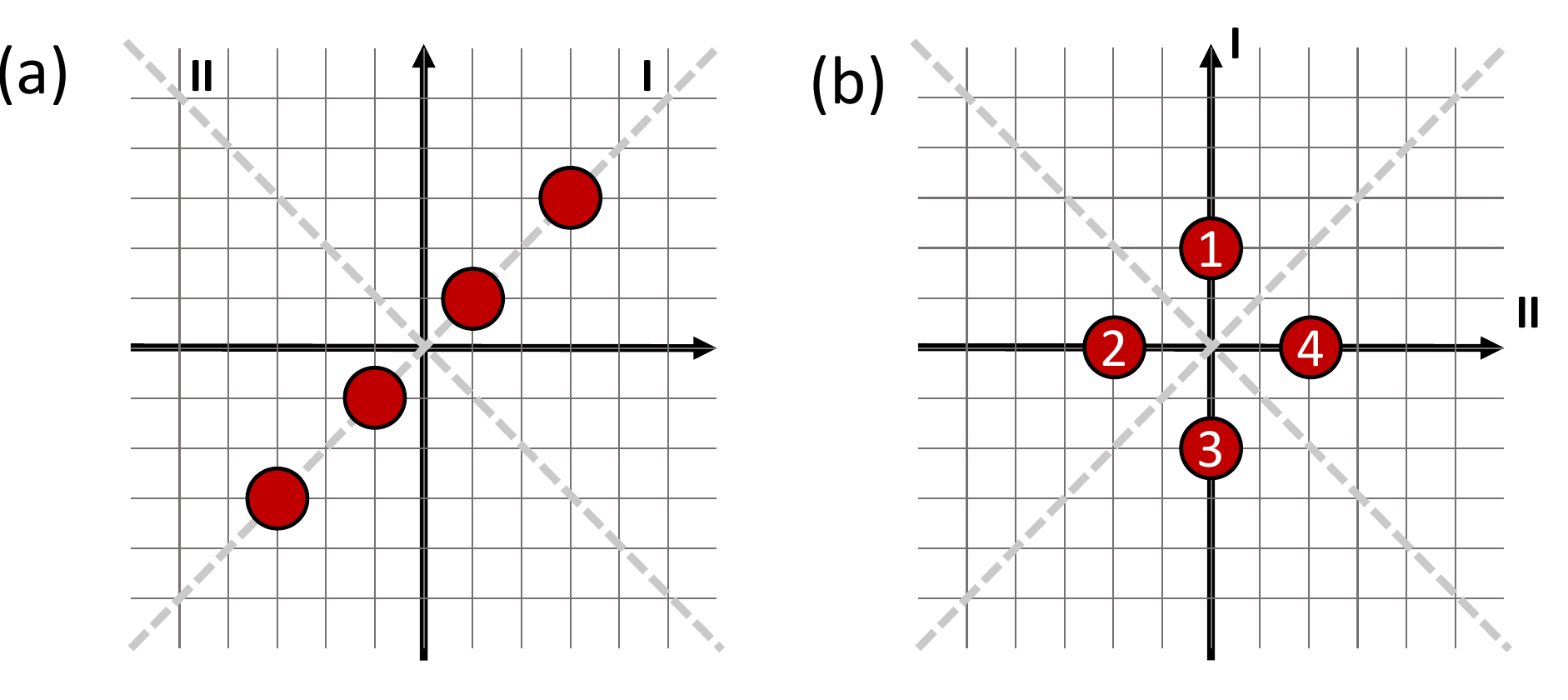}
\caption{Diagonal (left) and diamond (right) configurations of emitters studied in this work. 
The dashed grey lines indicate the symmetry axes that we use to aply our enhanced BL algorithm in each case.}
\label{configuration of emitters}
\end{figure}

\subsection{Diagonal configuration close to the band edge}
\label{sec:diagonal}

When the energy of the emitters is close to the band edges, $\Omega =\pm ( 4J - \epsilon)$, for $0 < \epsilon \ll J$, the emission is dominated by the bath modes at frequencies $\omega(\mathbf{k})=\pm (4J - \epsilon)$. Since near the band edges the density of states becomes almost flat, in the limit of $g/J \to 0$, the system exhibits Markovian dynamics.
Near the band edge, the bath modes have a long wavelength and the dispersion is almost isotropic, and thus in the Markovian limit, the system approaches the Dicke model where many emitters couple collectively to a single mode. Thus, one can observe collective radiance effects such as super- or subradiance.
Nevertheless, even at comparatively weak coupling, the dynamics deviates from the Markovian prediction, which we explore here.

To this end, we fix $\Omega/J=-3.95$ and consider a diagonal configuration along the $x=y$ line [labeled I in  Fig.~\ref{configuration of emitters}(a)] and symmetric with respect to the origin, i.e. pairs of emitters are placed in positions $(x_{\ell},x_{\ell})$ and $(-x_{\ell},-x_{\ell})$, with $x_{\ell}=2\ell-1$ for $\ell=1, \dots\, N_{e}/2$.

In this setup, the relevant symmetry transformations are reflections with respect to either diagonal.
As described in Sec.~\ref{sec: symmetries}, we can write the total Hamiltonian Eq.~\eqref{eq: Total Hamiltonian} in terms of bath modes with well-defined transformation under both reflections. In this basis the bath Hamiltonian results in a sum of four independent sectors $(s_{\mathrm{I}}=\pm,s_{\mathrm{II}}=\pm)$. 
Because the emitters lie on the main diagonal (I), there is no coupling to the bath sectors that transform antisymmetrically with respect to the $\mathcal{D}_{I}$ reflection ($s_{\mathrm{I}}=-$), which are thus irrelevant for the dynamics.
The interaction term, nevertheless, couples the emitters to both $s_{\mathrm{II}}=\pm$ sectors. It is easy to see that it can be written as a sum over the pairs of positions introduced above as
\begin{equation}
\op{H}_{\mathrm{int}}=\sum_{\ell=1}^{N_e/2} \left [O_1^{(\ell)}\otimes b_{(x_{\ell},x_{\ell})}^{(++)} +O_2^{(\ell)}\otimes b_{(x_{\ell},x_{\ell})}^{(+-)} + \mathrm{h.c.}\right ],
\label{eq:int_sym}
\end{equation}
where
\begin{equation}
b_{(x,x)}^{(+\pm)}=\frac{1}{\sqrt{2}}\left(a_{(x,x)}\pm a_{(-x,-x)} \right)
\end{equation}
are the annihilation operators for bath modes on the main diagonal with well-defined transformations, and
$O_{1,2}$ are linear combinations of operators acting on both emitters in the pair. It is convenient to define a collective basis for the pair as
\begin{align}
|0 \rangle_{\ell} &= |g\rangle_{x_{\ell}}\otimes |g \rangle_{-x_{\ell}}, \nonumber \\
|1+ \rangle_{\ell} &= \frac{1}{\sqrt{2}} \left( \sigma_{x_{\ell}}^++\sigma_{-x_{\ell}}^+\right)
|g\rangle_{x_{\ell}}\otimes |g \rangle_{-x_{\ell}}, \nonumber \\
|1- \rangle_{\ell} &= \frac{1}{\sqrt{2}} \left( \sigma_{x_{\ell}}^+-\sigma_{-x_{\ell}}^+\right)
|g\rangle_{x_{\ell}}\otimes |g \rangle_{-x_{\ell}} , \nonumber \\
|2\rangle_{\ell} &= |e\rangle_{x_{\ell}}\otimes |e \rangle_{-x_{\ell}},
\label{eq:coll_basis}
\end{align}
where we have used the subindex $x_{\ell}$ to identify the emitter in  position $(x_{\ell},x_{\ell})$.
States $\{ \ket{0},\,\ket{1+},\,\ket{2}\}_{\ell}$ span the subspace symmetric under $\mathcal{D}_{II}$ reflections, and $\{ \ket{1-}\}_{\ell}$ the antisymmetric one.
In this basis, the operators of the interaction term for each pair of emitters have the form
\begin{align}
O_1^{(\ell)}&=\frac{1}{\sqrt{2}} \left( \sigma_{x_{\ell}}^++\sigma_{-x_{\ell}}^+\right)= \left( \ket{1+}\bra{0}+\ket{2}\bra{1+} \right)_{\ell} \nonumber \\
O_2^{(\ell)}&= \frac{1}{\sqrt{2}} \left( \sigma_{x_{\ell}}^+ - \sigma_{-x_{\ell}}^+\right)=\left( \ket{1-}\bra{0}-\ket{2}\bra{1-}\right)_{\ell}.
\label{eq:ops_int_sym}
\end{align}

From the interaction term~\eqref{eq:int_sym} it is apparent that for each pair of emitters $\ell$ the only seed modes required for the BL are $b_{(x_{\ell},x_{\ell})}^{(++)}$ in the $(++)$ sector, and 
 $b_{(x_{\ell},x_{\ell})}^{(+-)}$ in the $(+-)$ sector. 
 Starting from the set of corresponding modes for all pairs we can proceed to apply the Block Lanczos procedure, as explained in Sec.~\ref{sec: Block Lanczos}, independently to the 
 bath Hamiltonian in each sector $H_{B}^{++}$ and $H_{B}^{+-}$, 
 resulting in the Block Lanczos Hamiltonians, $ \matr{H}_{BL}^{++}$ and $ \matr{H}_{BL}^{+-}$. 
 Since each of them can be truncated, as described in Sec.~\ref{sec: Block Lanczos}, at a different number of iterations, we obtain two ladders of width $N_e/2$ and lengths $L_{\mathrm{trunc}}^{++}$ and $L_{\mathrm{trunc}}^{+-}$. 
 For our numerical simulations, except when explicitly mentioned, we choose the same truncation $L_{\mathrm{trunc}}^{++}=L_{\mathrm{trunc}}^{+-}=L_{\mathrm{trunc}}$. 

 The interaction Hamiltonian couples each of the seed modes to an effective site for the emitters pair, with physical dimension 4, thus effectively connecting both ladders. 
 Fig.~\ref{fig:symmetric_chain} shows resulting geometry for the case of a single pair, with the emitter site in the center.
 
Using this representation we simulate the dynamics of the system with MPS for initial states of the form
\begin{equation}
\ket{\Psi(t=0)}=\ket{\psi_0}_{\mathrm{em}}\otimes \ket{\mathrm{vac}}_{\mathrm{bath}}.
\end{equation}

\begin{figure*}[t!]
\centering
    \includegraphics[width=0.95\linewidth]{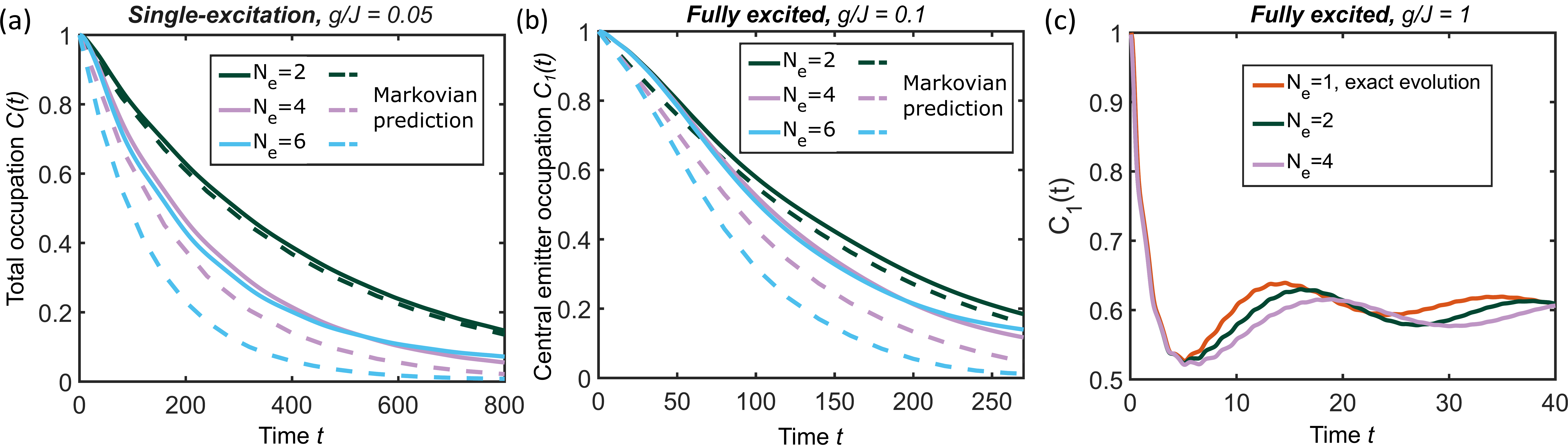}
\caption{
Evolution of occupation in the system of emitters for the diagonal configuration of Fig.~\ref{configuration of emitters}(a) for different coupling regimes, with emitters frequency near the band edge $\Omega/J=-3.95$ and effectively infinitely large bath. 
In (a), we show the weak-coupling single-excitation case, where we see perfect agreement between our algorithm and exact simulation. The Markovian prediction, however, already starts to fail for four and six emitters.
In (b), we consider the evolution from the fully excited initial state and plot the population of an emitter in the center of the chain (see \cref{eq: emitter occupation}). Again the Markovian prediction starts to deviate quite markedly for 4 and 6 emitters and collective radiance is suppressed due to the non-Markovian effects.
Finally, in (c) we show the strong coupling dynamics in otherwise the same setting as (b), where collective radiance is completely absent. However, we observe the appearance of a bound state.
Other parameters are as follows.
(a): $\{ \alpha, L_{\mathrm{trunc}}, \delta, D \}=\{ \{ 4, 1000, 0.4, 10\}, \{ 4, 600, 0.4, 10\}, \{ 4, 400, 0.4, 15\} \}$.
(b): $\{ \alpha, L_{\mathrm{trunc}}, \delta, D, n_{\mathrm{max}} \} = \{ 4, 600, 0.4, 10, 2  \}, \{ 4, 500, 0.2, 20, 2\}, \{ 4, 100, 0.4, 20, 3 \} \}$.
(c): $\{ L_{\mathrm{trunc}}, \delta, D, n_{\mathrm{max}} \} =\{ \{ 300, 0.05, 20, 2\}, \{100, 0.05, 20, 3 \} \}$. 
For figure (c) no truncation of the energy modes was applied.}

\label{omega395}
\end{figure*}

\subsubsection{Single-excitation case} 
  
We first analyze the case in which the state of the emitters contains a single excitation in a symmetric superposition.
Because the dimension of the single-excitation subspace is only linear in the total system size, we can solve this case exactly for large lattices, which provides us with a way to benchmark the algorithm.
The initial state of the emitters is then
\begin{align}
    \ket{\psi_0}_{\mathrm{em}}&=\frac{1}{\sqrt{N_{e}}}\sum_{i=1}^{N_e}\sigma_{i}^+ \ket{g}^{\otimes N_e}
    \nonumber \\
     & =\sqrt{\frac{2}{N_{e}}}\sum_{\ell=1}^{N_e/2}\ket{1+}_{\ell} \otimes_{p\neq \ell} \ket{0}_{p},
\label{eq: one excitation case-pairs}
     \end{align}  
     where the index $i$ in the first row runs over individual emitters, while indices $\ell$ and $p$ in the second row run over pairs.
    
    As observable, we measure the number of excitations in the system of emitters,
\begin{align}
    C(t)&=\sum_{i=1}^{N_{e}}
    \langle \Psi(t)|\sigma_i^{+}\sigma^-_i |\Psi(t)\rangle \nonumber \\
    &=\sum_{\ell=1}^{N_e/2} \bra{\Psi(t)} \Pi_{\ell}\ket{\Psi(t)},
    \label{eq: sum of emitter Occupation}
\end{align}
where the occupation of the $\ell$-th emitter pair can be written in the symmetric basis as
\begin{align}
\Pi_{\ell}&= \left( 2|2\rangle\langle 2|+|1+\rangle\langle 1+ |+|1- \rangle\langle 1-| \right)_{\ell}.    
\end{align}    
    
We consider a weak coupling value $g/J=0.05$, and simulate the dynamics for a bath lattice size $L_{B}=501\times 501$.
The results are shown in Fig.~\ref{omega395}(a). 
The MPS results coincide with those obtained from calculating the exact time evolution, with the use of the original Hamiltonian of \cref{eq: Total Hamiltonian} 
(before the Block Lanczos transformation), in the single-excitation subspace, thus verifying the results of our method. 
The figure shows a comparison of our numerics (solid lines) with the Markovian prediction (dashed lines) for 2, 4 and 6 emitters. 
Even though we are close to the band edges with the coupling $g/J$ being weak, the figure shows that non-Markovian effects are significant,
especially for larger number of emitters.
We heuristically identify this as the weak non-Markovian regime, since for the same $\Omega /J$ and in the limit of arbitrarily weak coupling $g/J$, the prediction of the Markovian master equation is recovered.

\subsubsection{Multiple excitation case} 

Whereas the single-excitation case serves as a benchmark, more interesting scenarios include initial states with multiple excitations, where 
exact diagonalization becomes inefficient and our method provides an advantage.
To explore this situation, we consider the initial state where all emitters are initially excited, namely
\begin{align}
    \ket{\psi_0}_{\mathrm{em}} &= \ket{e}^{\otimes N_e}
 =\ket{2}^{\otimes N_e/2}.
\label{eq: many excitation case-pairs}
     \end{align}  

As observable, we focus on the average occupation of the innermost pair of emitters ($\ell=1$), namely
\begin{align}
\label{eq: emitter occupation}
C_{1}(t)=\frac{1}{2}\langle \Psi(t)|\Pi_{1} |\Psi(t)\rangle.
\end{align}
For a weak value of the coupling $g/J=0.1$
it is reasonable to compare the numerical results of our (quasi-exact) simulations to those of the Markovian approximation.
Fig.~\ref{omega395}(b) shows this comparison for a lattice of size $L_{B}=501 \times 501$.
Again we observe substantial deviation of the MPS numerics (solid) from the Markovian prediction (dashed lines) already at early times, more significant for larger number of emitters.
This shows that the non-Markovian effects prevent collective radiation from occurring as efficiently as in the Markovian setting.

We also consider a much stronger coupling $g/J=1$ for a system size $L_{B}=301\times 301$. In this regime, the system is strongly non-Markovian, and we find signals of a bound state, as predicted in Ref.~\cite{Shi2016, gonzalez20172Dreservoirs, gonzalez2017markovian}. Specifically, Ref.~\cite{Shi2016, gonzalez20172Dreservoirs, gonzalez2017markovian} discuss the existence of a single-excitation bound state, and Ref.~\cite{Shi2016} the existence of many-excitation bound states in two-level systems that are coupled to bosonic baths. Importantly, for two-dimensional tight-binding baths there always exists a single-excitation bound state independently of the parameter regime~\cite{Shi2016}. Additionally, the existence of many-excitation bound states is guaranteed analytically when the coupling $g/J$ is either very strong or very weak~\cite{Shi2016}.

\Cref{omega395}(c) shows the results for 1, 2 and 4 emitters.
Different from the \textit{weak non-Markovian case}, in this case the occupation $C_1(t)$ does not decay to zero, reflecting the presence of the bound state.
Interestingly, as the number of emitters (and excitations) increases, the frequency of the oscillations decreases, and effect opposite in nature to collective radiance, which proceeds faster with more emitters. 
We note here that the times reached are shorter than in previous cases, partly because the strong emitter--bath coupling is more challenging to simulate, and partly because the relevant dynamics also proceeds faster.
This regime is numerically more challenging, since strong coupling (i) means we need a larger maximal occupation number of bath modes $n_{\max}$, and (ii) implies that it is no longer possible to truncate the bath in energy space. Moreover, and closely related, we have faster dynamics, which (iii)  leads to a faster increase of entanglement, and (iv) forces us to choose smaller Trotter steps.
Despite these challenges, our method allows us to simulate the main physical effect: the dynamical preparation of an excited bound state.


\subsection{Diamond configuration in the middle of the band: bound state}
\label{sec:diamond}

\begin{figure}
\centering
    \includegraphics[width=0.85\linewidth]{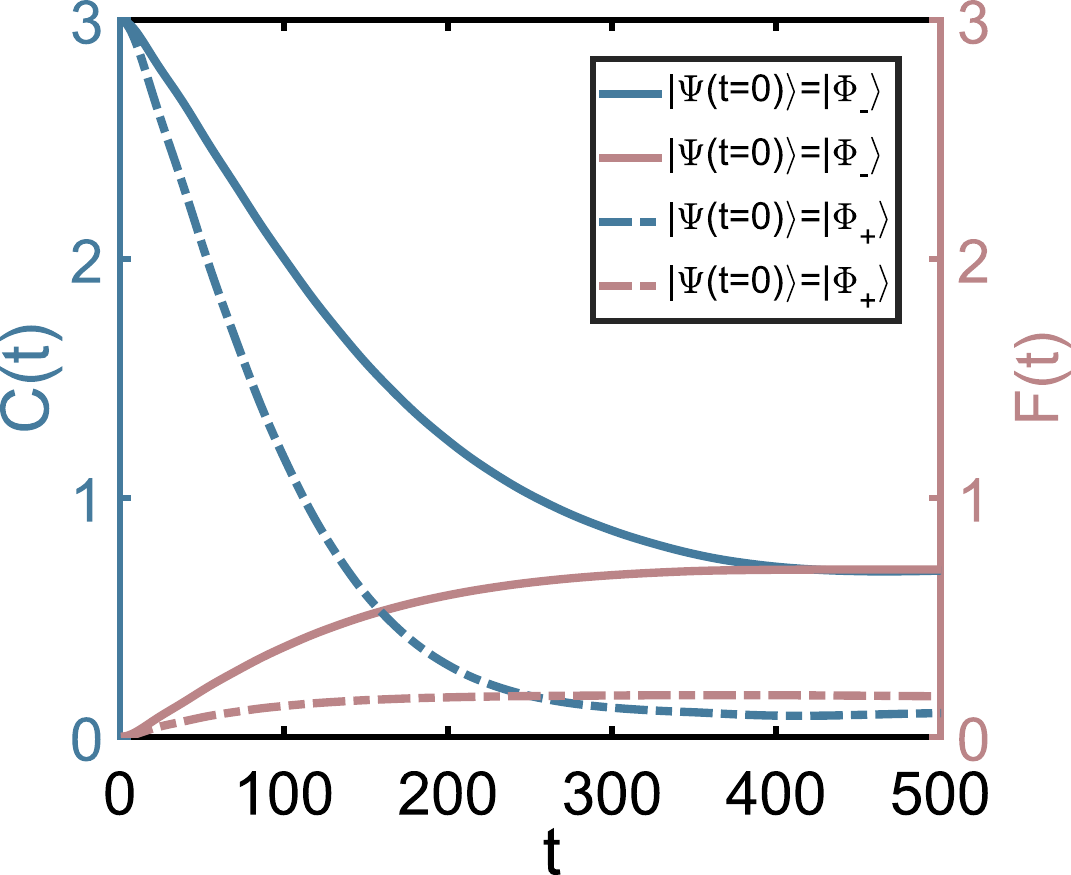}
\caption{The emitter population (\cref{eq: sum of emitter Occupation}, blue lines), and the fidelity of the bound state with the initial state of the system (\cref{eq: fidelity}, pink lines). We observe that the probability of successfully preparing the system in a bound state is much better if the initial state observes the same symmetries as the single-excitation bound state.
The four emitters are prepared in a diamond configuration of Fig.~\ref{configuration of emitters}(b), with the initial states of Eq.~\eqref{eq: antisymmetric initial state 3 exc} and Eq.~\eqref{eq: symmetric initial state 3 exc, pairs}. The detuning is $\Omega=0$ and the coupling $g/J=0.05$ . The parameters used for the TN simulations are the following $\{ L_{\mathrm{trunc}}^{++}, L_{\mathrm{trunc}}^{+-}, \alpha, \delta, D, n_{\mathrm{max}} \} = \{ 700, 75, 4, 0.7, 15, 2  \}$.
}
\label{Bound State}
\end{figure}

We now consider the situation in which the emitters frequencies lie in the middle of the band ($\Omega=0$).
In this case, the system is non-Markovian for all values of the coupling $g/J$~\cite{LANGLEY1996447, mekis1999lasing}, because the density of states is divergent.
The single-excitation physics of this model was studied in detail in~\cite{gonzalez20172Dreservoirs, gonzalez2017markovian}, which found that for $\Omega=0$, the emission into the bath is directional, dominated by the two quasi-1D orthogonal modes $k_{x}\pm k_{y}=\pi$.
For four emitters placed in a diamond configuration around the origin, as depicted in \cref{configuration of emitters}(b), a bound state arises, and an excitation can get trapped between the emitters.
The exact form of this bound state is not known analytically, but it has a significant overlap with the one-excitation antisymmetric state~\cite{gonzalez2017markovian}

\begin{align}
    |\phi_{\mathrm{em}}\rangle &=\frac{1}{2} (\sigma^{+}_{1}-\sigma^{+}_{2}+\sigma^{+}_{3}-\sigma^{+}_{4})|g\rangle^{\otimes 4}, 
    \label{eq: one excitation antisymmetric state}
\end{align}
where the order of the emitters is defined in Fig.~\ref{configuration of emitters}(b).

\begin{figure}
\centering
    \includegraphics[width=0.85\linewidth]{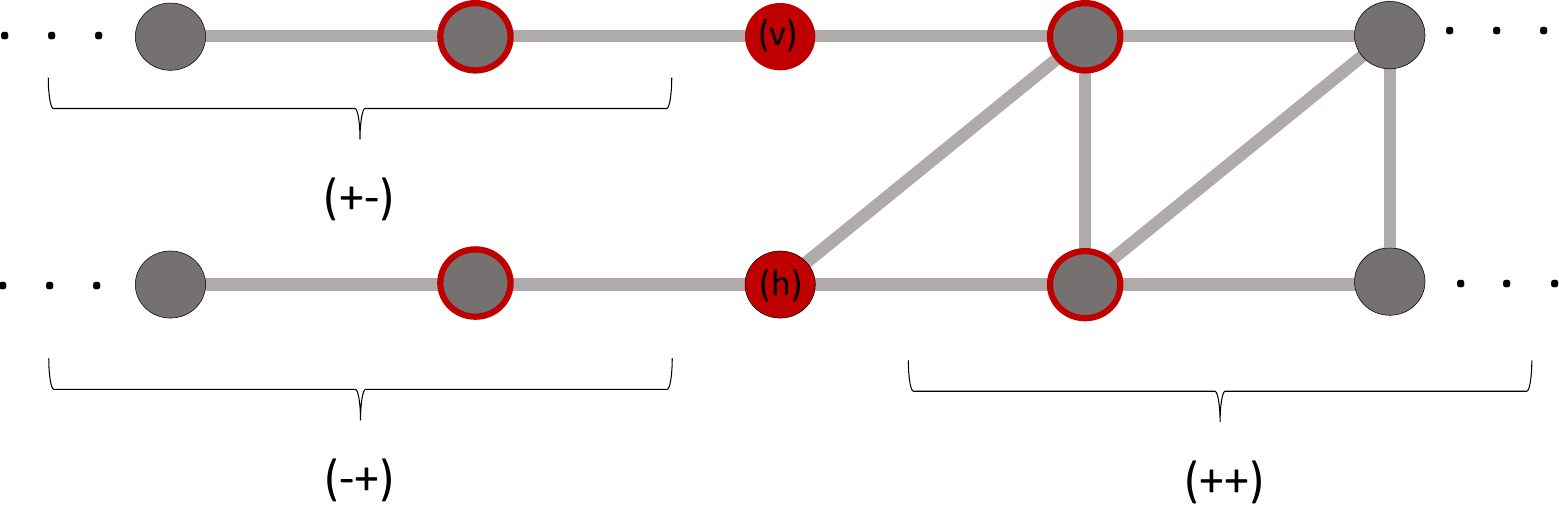}
\caption{
    In the case of the diamond configuration of Fig.~\ref{configuration of emitters}(b), two effective emitter sites (red circles), corresponding to the vertical and horizontal pair, couple to three of the four symmetry subsectors of the bath. The whole Hamiltonian can be rearranged in a 2-leg ladder as shown in the figure.
}
\label{fig: diamond and symmetries}
\end{figure}

We focus on the case of this diamond configuration, with four emitters located in positions $(\pm2,0)$ and $(0,\pm2)$.
In this case, the system preserves the original $D_{4}$ symmetry of the lattice, but we can also use a reduced symmetry, 
considering only reflections with respect to the horizontal and vertical axes. 
This allows us to use a very similar formalism as in the diagonal configuration.
Writing the Hamiltonian in terms of bath modes with well-defined transformation under these reflections, we again obtain four 
independent bath sectors. 
But, different to the diagonal configuration, now three different sectors couple to the emitters,
which are now grouped in two effective sites, one for the horizontal pair and one for the vertical one.
The form of the interaction Hamiltonian is
\begin{align}
\op{H}_{\mathrm{int}}&= O_1^{(\mathrm{v})}\otimes b_{(0,2)}^{(++)} +O_2^{(\mathrm{v})}\otimes b_{(0,2)}^{(+-)} \nonumber \\
&+ O_1^{(\mathrm{h})}\otimes b_{(2,0)}^{(++)} +O_2^{(\mathrm{h})}\otimes b_{(2,0)}^{(-+)}  + \mathrm{h.c.},
\label{eq:int_sym_diamond}
\end{align}
where in this case the $(s_1,s_2)$ labels identifying the modes respectively refer to the reflections with respect to horizontal and vertical axes. 
The corresponding symmetric bath modes are
\begin{align}
b_{(0,2)}^{(+\pm)}=\frac{1}{2}\left( a_{(0,2)}\pm a_{(0,-2)} \right), \nonumber \\
b_{(2,0)}^{(\pm+)}=\frac{1}{2}\left( a_{(2,0)}\pm a_{(-2,0)} \right).
\end{align}
The emitters operators have the same expression as in Eq.~\eqref{eq:ops_int_sym}, now referring to the vertical (v) and horizontal (h) emitter pair.
Notice that now both pairs couple to the (++) bath sector. The BL transformation of this sector, starting from modes $b_{(0,2)}^{(++)}$ and $b_{(2,0)}^{(++)}$,
yields a 2-wide ladder. On the other hand, sectors (+-) and (-+) couple only to one pair, each, so each of the corresponding BL transformations produces 
a simple chain. The total Hamiltonian is then a ladder of width 2, with the effective emitters sites in the central rung and both legs connected only in the (++) half (see Fig.~\ref{fig: diamond and symmetries}).

In order to go beyond the single-excitation case studied in  Ref.~\cite{gonzalez20172Dreservoirs}, we consider an initial state of the emitters containing
 three excitations, with the same symmetries as the single-excitation state of Eq.~\eqref{eq: one excitation antisymmetric state},
\begin{align}
    \ket{\Phi_-}&=\frac{1}{2} (\sigma^{-}_{1}-\sigma^{-}_{2}+\sigma^{-}_{3}-\sigma^{-}_{4})|e,e,e,e\rangle \nonumber \\
&=
\frac{1}{\sqrt{2} } \left( |1+ \rangle_{\mathrm{v}} \otimes |2 \rangle_{\mathrm{h}} - |2 \rangle_{\mathrm{v}}  \otimes |1+ \rangle_{\mathrm{h}} \right).
    \label{eq: antisymmetric initial state 3 exc}
\end{align}

We simulate the dynamics of this system with the MPS ansatz for a system size $L_{B}=701\times 701$.
Fig.~\ref{Bound State} shows the time dependence of the emitter occupation (blue) of Eq.~$\eqref{eq: sum of emitter Occupation}$. 
The convergence to a non-vanishing value is a signature of the bound state.
To verify this, we compute the fidelity between the reduced state of the emitters during the evolution and the single-excitation ansatz for the bound state Eq.~\eqref{eq: one excitation antisymmetric state}
\begin{align}
\label{eq: fidelity}
  F(\rho, \phi_\mathrm{em})= \langle \phi_{\mathrm{em}}  |\rho(t)| \phi_{\mathrm{em}} \rangle.
\end{align}
Here, $\rho(t)$ is the reduced density matrices of the the emitters at time $t$.
The result, shown as a pink line in  Fig.~\ref{Bound State}, indicates that the fidelity increases in time, and converges to a finite value $\sim 0.7$, which signals the dynamical preparation of an excitation in the bound state.

For comparison, we consider a different initial state
\begin{align}
    \ket{\Phi_+}&= \frac{1}{\sqrt{2} } \left( |1+ \rangle_{\mathrm{v}} \otimes |2 \rangle_{\mathrm{h}} + |2 \rangle_{\mathrm{v}}  \otimes |1+ \rangle_{\mathrm{h}} \right),
    \label{eq: symmetric initial state 3 exc, pairs}
\end{align}
which is orthogonal to $ |\Phi_{-}\rangle$.
Interestingly, even though $|\Phi_{+}\rangle$ has different symmetries than the single-excitation bound state ansatz $\ket{\phi_\mathrm{em}}$ [\cref{eq: one excitation antisymmetric state}], the fidelity of the time-evolved state with the bound state still becomes finite and saturates to a low but non-zero value.
This is possible because only the symmetry of the total system, and not that of the emitters independently, is preserved during the evolution.

\section{Discussion}\label{sec: Discussion}

We propose a method to study a system of quantum emitters coupled linearly to a non-interacting bath.
Our method is based on a Block Lanczos transformation of the bath that maps the original problem onto a strip geometry.
We simulate this quasi-one-dimensional geometry (quasi-exactly) using tensor network methods.
While this mapping is in principle possible for any such system, the cost of the simulation increases fast with the number of emitters. 
Thus, a key contribution of this work is  a number of techniques that substantially simplify the simulations.
To treat systems with many emitters more efficiently, we show that one can exploit spatial symmetries of the system.
Since our simulation method is built on trotterization, another regime that becomes computationally expensive is that of close-to-Markovian systems with fast bath dynamics and comparatively slow system dynamics. We show that this can be alleviated by an effective truncation of the bath in energy space. 

Using this method we study the time evolution of a system of multiple two-level quantum emitters coupled to a square two-dimensional bosonic bath, in weak and strong coupling regimes. We verify our calculations against exact results obtained in the single-excitation and two-excitation subspace, and go beyond previous studies by exploring multi-emitter, multiple excitation setups, looking in particular at collective radiance and the formation of bound states.

To explore the effect of non-Markovianity on collective radiance, we consider the regime where the emitter energies are placed close to the band edge, such that they couple to long-wavelength modes with almost isotropic dispersion. In the limit of weak coupling and close to the band edge, the system would become a Dicke model. Using our method, we explore how non-Markovian effects destroy the collective radiance.
We observe the deviation from the Markovian prediction to grow both with increasing coupling, and increasing emitter number.
In the middle of the band, the square lattice has a divergent density of states, which makes the system non-Markovian even for arbitrarily small coupling. In this fundamentally non-Markovian regime, we study how a single-excitation bound state can dynamically form from a many-excitation initial state.

Beyond the cases we study in this work, the method can be easily extended to any non-interacting bath, in any spatial dimension, also those of fermionic character. 
Even though we have only simulated scenarios where the bath is initialized in the vacuum, it is equally possible to treat arbitrary initial states, including thermal ensembles, as done in simple chain mappings~\cite{Prior2010,devega2015thermofield,Schwarz2018}.
Although we have considered real time evolution in this work, in principle the same technique can be used to study equilibrium problems in condensed matter impurity systems, e.g., through DMRG or imaginary time evolution. This could then be applied in dynamical mean-field theory~\cite{Georges1996,Ganahl2015,Wolf2015,Bauernfeind2017}.
Finally, it would also be interesting to explore other tensor network ansatzes to simulate the time evolution in the strip geometry or to use this mapping to simulate multi-emitter systems on quantum computers.

\begin{acknowledgments}
This work was partly supported by  the DFG (German Research Foundation) under Germany's Excellence Strategy -- EXC-2111 -- 390814868,  Research Unit FOR 5522 (grant nr. 499180199), and TRR 360 - 492547816. Plus.
D.\ M.\ acknowledges support from the Novo Nordisk Foundation under grant numbers NNF22OC0071934 and NNF20OC0059939.
\end{acknowledgments}

\bibliography{main.bib}
\end{document}